# Symmetry-guided modal control in elliptical femtosecond-laser-written photonic waveguides

Tadas Paulauskas[1]*, Ubaid Ur Rehman[1], Eimantas Dermauskas[2], and Valdemar Stankevič[2]

[1]State Research Institute Center for Physical Sciences and Technology, Department of Optoelectronics, Vilnius LT-10257, Lithuania

[2]State Research Institute Center for Physical Sciences and Technology, Department of Laser Technologies, Vilnius LT-10257, Lithuania

*tadas.paulauskas@ftmc.lt

**Abstract:** Few-mode photonic circuits can increase functionality without multiplying waveguide paths, but bends and fabrication errors mix their transverse modes. We investigate a strategy in which waveguide confinement and perturbation parity are engineered together in vertically elliptical femtosecond-laser-written glass waveguides. The intended basis comprises the even 1S mode and the vertically odd $2P_y$ mode. No window-converged $2P_x$ state is resolved for a lower-confinement (LC) design, whereas a higher-confinement (HC) design guides $2P_x$ and must isolate it by symmetry. In scalar beam-propagation calculations, the 1S–$2P_y$ propagation-constant splitting predicts the optimized periods of a vertically modulated coherent modal splitter to within 1.0%. Horizontal S-bends remain parity mismatched to 1S↔$2P_y$, while the symmetry-allowed HC 1S→$2P_x$ transfer reaches only 0.6% at the largest displacement. The HC design retains approximately the same $2P_y$ power at 150 μm displacement that the LC design retains at 40 μm. Thermal and stochastic writing-error calculations expose the resulting trade-off: stronger confinement improves modal-power retention, but $x$-parity-breaking writing jitter can populate guided $2P_x$. These results show how modal-basis engineering can shift part of the crosstalk-control burden from the trajectory to waveguide symmetry, supporting joint path–mode degrees of freedom in quantum photonic applications.

## 1. Introduction

Femtosecond-laser writing (FLW) enables three-dimensional routing of buried waveguides in transparent dielectrics, making it a flexible platform for volumetric photonic integrated circuits [1,2]. This capability has enabled single-mode circuits, multimode-interference devices, mode-selective couplers, quantum photonic circuits, and reconfigurable glass photonic processors [3–7].

Most FLW circuits use single-mode path encoding, leaving transverse spatial modes largely unused. A controlled few-mode basis could support modal beam splitters, interferometers, mode-division interfaces, and compact path–mode processors. The difficulty is not simply generating a higher-order mode: nearly degenerate $LP_{11}$-like orientations are susceptible to routing and fabrication perturbations. Established solutions typically optimize a coupling element or trajectory—for example, mode-selective couplers and long-period gratings [4,5] or optimized bends [6]—after the supported modal manifold has been chosen. Elliptical laser-written guides offer another degree of freedom because their anisotropy lifts the first-order-mode degeneracy [7,8]. This motivates a complementary question: can the modal basis itself be chosen so that routine routing and deliberate conversion address orthogonal symmetry channels?

We address this question by engineering a vertically elongated core whose intended basis is the even 1S mode and the vertically odd $2P_y$ mode. In the lower-confinement (LC) design, no $2P_x$ trial passes the transverse-window convergence test, so the unwanted orientation is excluded through weak guidance or radiation. In the higher-confinement (HC) design, $2P_x$ is a window-converged guided mode but remains outside the intended dynamics under vertically odd control. Comparing these regimes exposes a trade-off: higher confinement improves retention of the extended $2P_y$ mode, but creates a resolved leakage channel when laser-writing errors break $x$-parity.

The design rule is to co-engineer the supported modal basis and the symmetry of the perturbation acting on it. A horizontal trajectory perturbation is odd in $x$ and therefore parity mismatched to 1S↔$2P_y$ at first order, whereas a periodic vertical displacement is odd in $y$ and can be phase-matched to that transition. The contribution is thus not the sinusoidal trajectory alone, but the use of modal-spectrum engineering and perturbation symmetry as coupled design variables. The LC/HC comparison separates guidance-based exclusion from symmetry-based selectivity.

We test this rule in progressively more practical settings. After validating the 640 nm modal bases by transverse-window convergence, we examine reciprocal modal splitting and sensitivity to waveguide and trajectory parameters. We then use horizontal S-bends as the complementary parity test, impose a surface-heater-like perturbation, and compare parity-preserving with parity-breaking stochastic writing errors. An illustrative figure places these elements in a conceptual path–mode circuit. All calculations use a scalar, selected-polarization model, while birefringence and device-specific thermal transport remain experimental calibration requirements.

## 2. Principle of symmetry-selective modal coupling

### 2.1 Parity selection in an engineered modal basis

For weakly guiding waveguides at fixed polarization, the symmetry argument follows directly from standard scalar coupled-mode theory [9,10]. We apply it to the slowly varying scalar envelope. Throughout the numerical results, $\beta$ denotes the envelope propagation constant relative to the carrier reference, $k_{ref} = k_0 n_{clad}$. The absolute propagation constant is $k_{ref} + \beta$, and only differences between modal $\beta$ values enter the phase-matching condition. We expand the envelope field in eigenmodes of the unperturbed guide as

(1) $$E(x, y, z) = \sum_m a_m(z)\, \psi_m(x, y) \exp(i\beta_m z),$$

where $\psi_m(x, y)$ is the transverse mode field, $\beta_m$ is the cladding-referenced propagation-constant offset, and $a_m(z)$ is a slowly varying modal amplitude. To first order, an index perturbation $\Delta n(x, y, z)$ couples two modes through an overlap of the form:

(2) $$\kappa_{nm}(z) \propto \iint \psi_n^*(x,y)\, \Delta n(x,y,z)\, \psi_m(x,y)\, dx\, dy.$$

The mode parities make the design choice explicit. The fundamental 1S mode is even in *x* and *y*, whereas $2P_y$ is even in *x* and odd in *y*. The orthogonal first-order mode $2P_x$ is odd in *x* and even in *y*. These labels describe nodal parity and orientation rather than exact step-index fiber LP modes. The intended subspace is $\{1\mathrm{S}, 2P_y\}$. For a small horizontal displacement of an otherwise mirror-symmetric unperturbed index profile, a first-order expansion gives the perturbation below

(3) $$\Delta n_x(x,y,z) \simeq -\delta x(z)\partial_x\, n_0(x,y).$$

The transverse derivative of the unperturbed index is odd in *x* and even in *y*. The resulting perturbation is symmetry-mismatched to the $1S \leftrightarrow 2P_y$ transition, so the overlap integral vanishes:

(4) $$\kappa_{S,P_y}^{(x)} \propto \iint \psi_S(x,y)\, x\, \psi_{P_y}(x,y)\, dx\, dy = 0.$$

By contrast, the same perturbation would be symmetry matched to the horizontally odd transition $1S \leftrightarrow 2P_x$, meaning $\kappa_{S,P_x}^{(x)} \neq 0$. This is the reason for excluding or detuning $2P_x$. The remaining $1S/2P_y$ subsystem is therefore decoupled at first order from the dominant horizontal perturbation.

### 2.2 Vertical perturbation as the control channel

Intentional modal exchange requires the complementary perturbation symmetry. For a small vertical displacement, the corresponding first-order index perturbation is

(5) $$\Delta n_y(x,y,z) \simeq -\delta y(z)\partial_y\, n_0(x,y).$$

The transverse derivative of the unperturbed index is even in *x* and odd in *y*. The resulting perturbation is parity matched to 1S↔$2P_y$, giving $\kappa_{S,P_y}^{(y)} \neq 0$. If the vertical perturbation is periodic, efficient coherent exchange also requires its spatial frequency to compensate the propagation-constant mismatch,

(6) $$K \simeq \Delta\beta_{S,P_y} = \beta_S - \beta_{P_y},\ \Lambda \simeq \frac{2\pi}{|K|}.$$

The trajectory period therefore sets the longitudinal phase matching, its amplitude sets the accumulated coupling strength, and the elliptical cross-section determines both the transverse selection rule and the propagation-constant splitting. The vertically modulated section can consequently be viewed as a resonantly driven two-level modal system.

This parity argument concerns scalar spatial modes. FLW glass waveguides can also exhibit stress-induced or form birefringence [11,12], so a vector treatment assigns polarization-dependent propagation constants to each spatial mode and makes the phase-matching condition polarization specific,

$$(7) \qquad \Lambda_q \simeq \frac{2\pi}{|\beta_{S,q} - \beta_{P_y,q}|}, q \in \{H, V\}.$$

The calculations should therefore be interpreted for a selected polarization channel, or for a fabrication regime in which birefringence is small relative to the 1S– $2P_y$ propagation-constant separation. In experiment, the input polarization and modal spectrum should be measured before the modulation period is calibrated. Simultaneous operation for both polarizations would require low-birefringence writing or separate tuning of their resonances.

## 3. Validating the elliptical modal basis

### 3.1 Numerical methods

We used a custom scalar split-step Fourier beam-propagation method (BPM) for low-contrast glass waveguides. Device calculations were performed at $\lambda$ = 640 nm with $n_{clad}$ = 1.507. The LC design used $\Delta n_0$ = 2.5 × 10⁻³, $w_x$ = 4.0 µm, and ellipticity $w_y/w_x$ = 1.60; the HC design used $\Delta n_0$ = 3.0 × 10⁻³, $w_x$ = 4.5 µm, and the same ellipticity. Routine propagation used an 80 µm × 80 µm transverse window on a $512^2$ grid with $\Delta z$ = 2 µm and absorbing boundary layers. The $2P_y$ thermal convergence check was repeated on a 120 µm × 120 µm, $768^2$ grid at the same transverse sampling.

The written index modification was represented by an elliptical Gaussian profile with the stated horizontal and vertical full widths at half maximum (FWHM). Straight-guide fields were obtained by β-filtered propagation, normalization, and Gram-matrix checks. A mode was accepted for device projections only after its field and propagation constant converged between the 80 and 120 µm windows. We refer to these window-converged fields as the validated modal basis. All powers are normalized to launched power: $P_m = |\langle\psi_m|E\rangle|^2$ is a resolved modal power, $P_{out}$ is total output, $P_{out} - \sum_m P_m$ is unresolved in-window power, and $1 - P_{out}$ is absorber-removed power.

### 3.2 Modal design window and convergence

The design maps in figure 1 locate parameter regions in which a vertically elliptical guide supports the intended 1S/$2P_y$ basis. The 640 and 810 nm maps show how ellipticity, core width, and index contrast separate the first-order branches, and the remainder of the paper uses two independently validated 640 nm designs. These realize different isolation mechanisms: guidance-based $2P_x$ exclusion in LC and symmetry-based exclusion in HC, where $2P_x$ is guided.

Transverse-window convergence distinguishes guided modes from computational-box states. Across all accepted LC and HC modes, cropped field-overlap powers exceed 99.996% and $|\Delta\beta|$ remains below 0.25 rad m⁻¹ when the window is enlarged from 80 µm/$512^2$ to 120 µm/$768^2$ at fixed sampling. The LC 1S and $2P_y$ states pass this test, but no LC $2P_x$ trial does. All three HC states pass, confirming that HC $2P_x$ is a guided scalar mode rather than a box artifact.

This numerical validation sets up the design trade-off used throughout the paper. The more extended LC $2P_y$ mode is expected to radiate more readily under bending and heating, but the absence of a window-converged $2P_x$ channel provides guidance-based isolation. Higher confinement improves retention, while making fidelity depend on preserving $x$-parity. The relevant claim is therefore first-order symmetry selectivity, not immunity to arbitrary perturbations.

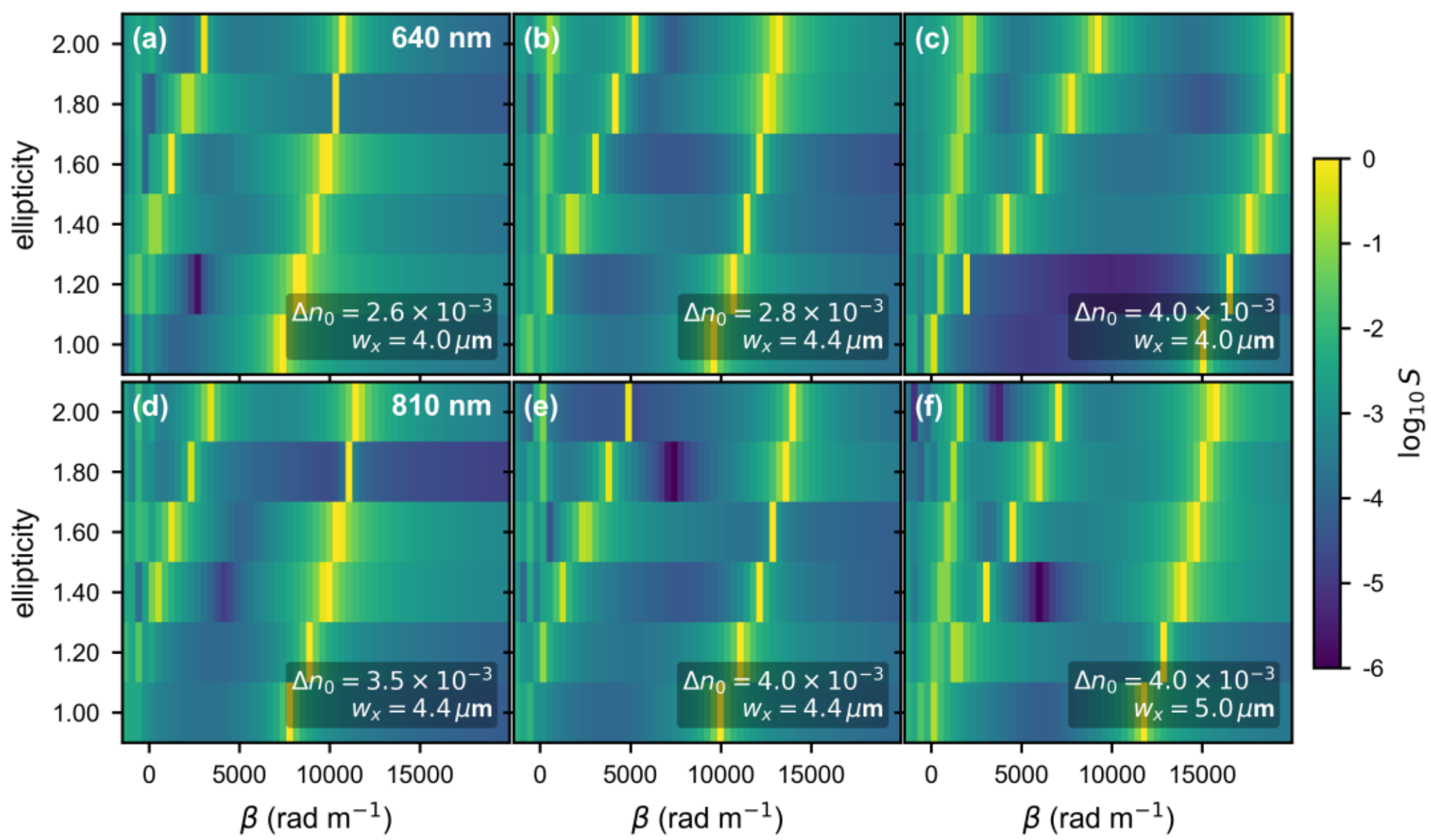

**Figure 1.** Modal design maps for vertically elliptical weakly guiding waveguides. Spectral density is plotted against the fitted, cladding-referenced envelope propagation constant $\beta$ and ellipticity for selected peak index contrasts $\Delta n_0$ and horizontal widths $w_x$. Panels (a–c) are calculated at 640 nm and panels (d–f) at 810 nm. Positive-$\beta$ branches identify increasingly confined modal states.

Figure 2 makes this distinction concrete at 640 nm. The LC projection basis contains validated 1S and $2P_y$ modes, whereas the HC basis also contains a weakly bound but window-converged $2P_x$ mode. The fitted, cladding-referenced propagation-constant offsets are $\beta_{1S} = 9.19 \times 10^3$ rad m$^{-1}$ and $\beta_{2P_y} = 8.29 \times 10^2$ rad m$^{-1}$ for LC, and $\beta_{1S} = 1.37 \times 10^4$ rad m$^{-1}$, $\beta_{2P_y} = 4.03 \times 10^3$ rad m$^{-1}$, and $\beta_{2P_x} = 6.73 \times 10^2$ rad m$^{-1}$ for HC. The corresponding 1S–$2P_y$ beat periods are 0.752 and 0.647 mm, providing parameter-free starting points for the driven modal splitters.

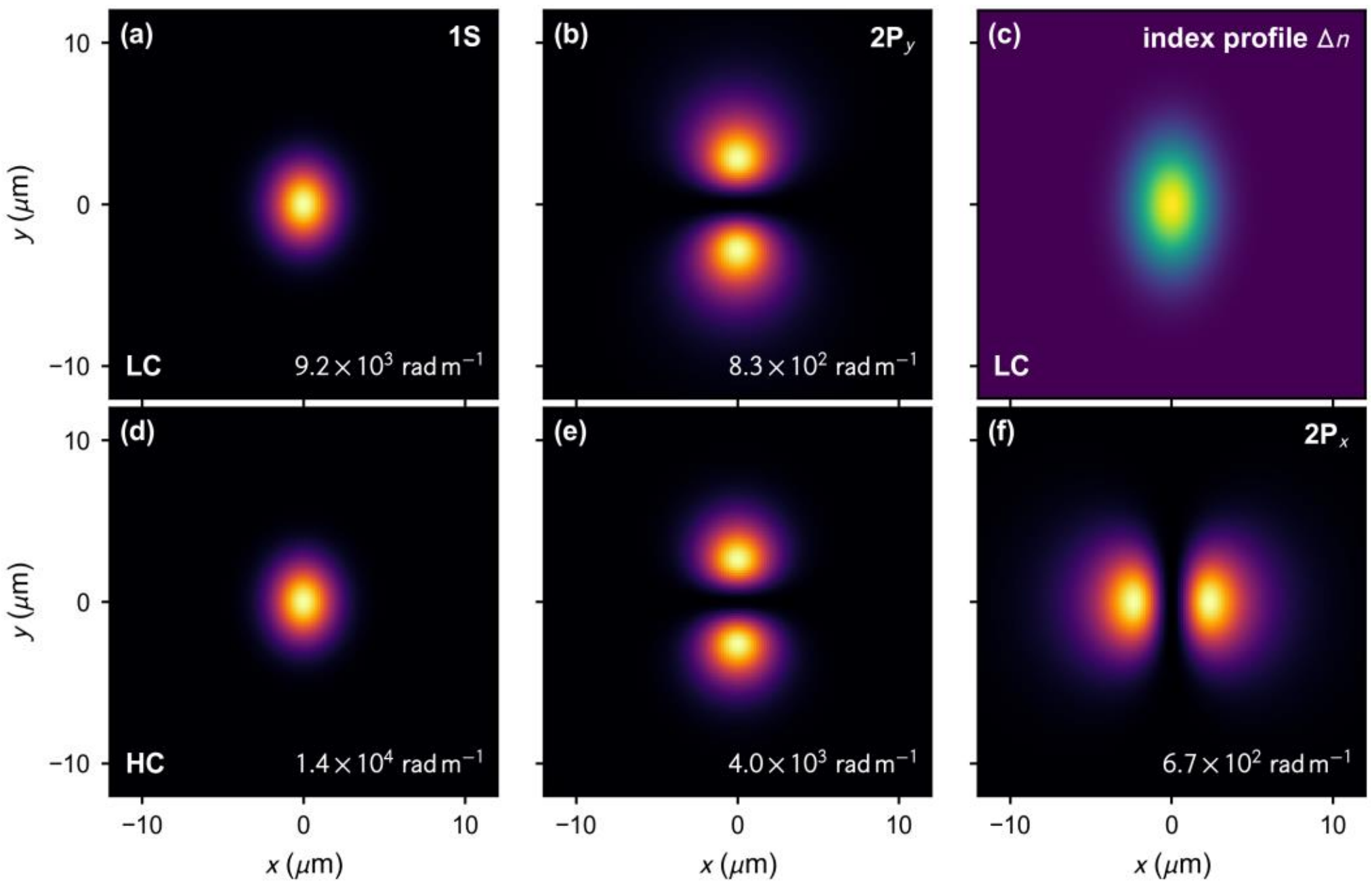

**Figure 2.** Validated 640 nm modal bases. The upper row shows the lower-confinement (LC) design: normalized 1S and $2P_y$ intensities and the corresponding index profile ($\Delta n_0$ = 2.5 × 10⁻³, $w_x$ = 4.0 µm, ellipticity = 1.60). The lower row shows the higher-confinement (HC) design: normalized 1S, $2P_y$, and $2P_x$ intensities ($\Delta n_0$ = 3.0 × 10⁻³, $w_x$ = 4.5 µm, ellipticity = 1.60). Annotations give the cladding-referenced $\beta$ values in rad m⁻¹. Each modal intensity is normalized independently.

## 4. Vertical trajectory modulation as a coherent modal splitter

### 4.1 Design and phase matching

Having established the modal basis, we next test whether a perturbation of the correct parity can drive controlled exchange without populating HC *2Pₓ*. We implement the control as a finite sinusoidal vertical centerline displacement, referred to in the figures as the wobbler. The induced first-order index perturbation is odd in $y$ and even in $x$, and is therefore matched to 1S↔$2P_y$ but symmetry mismatched to $2P_x$. Both validated input modes are launched to test reciprocity and to separate conversion from input-dependent attenuation. In the full-amplitude region, the centerline is

$$y_c(z) = A\cos\left(\frac{2\pi z}{\Lambda}+\phi\right). \tag{8}$$

The LC splitter uses $A$ = 0.247 µm and $\Lambda$ = 0.744 mm, while the HC splitter uses $A$ = 0.220 µm and $\Lambda$ = 0.644 mm. Equation (8) describes the full-amplitude region, with initial trajectory phase $\phi$ = 0 in the reported calculations. Each 10 mm device contains four full-amplitude periods joined to the straight guide by 0.4 mm cosine ramps. The shorter HC period follows from its larger 1S–$2P_y$ propagation-constant separation.

The independently extracted modal splittings give coupled mode theory phase-matching periods $\Lambda$ = 0.752 mm for LC and 0.647 mm for HC. The optimized BPM values, 0.744 and 0.644 mm, differ by 1.0% and 0.5%, respectively. This agreement connects the device response directly to the modal basis rather than treating the optimized trajectory purely as an empirical fit.

At these settings, a 1S launch produces near-equal absolute endpoint powers in both designs: $P_{1S}$ = 0.495 and $P_{2P_y}$ = 0.490 for LC, and 0.492 and 0.490 for HC (figure 3a). The corresponding powers contained in the validated pair are 0.985 and 0.982, while HC $2P_x$ remains below $10^{-12}$. The near-equal split therefore occurs within the intended pair without activating the symmetry-mismatched HC channel.

Reversing the launched mode reproduces the cross-converted power to within $2 \times 10^{-5}$ in the numerical model, consistent with reciprocal 1S↔$2P_y$ exchange at the numerical resolution. Reciprocity of the converted power does not imply equal total transmission: total output for $2P_y$ launch is 0.951 (LC) and 0.939 (HC), compared with 0.989 and 0.986 for 1S launch. Equal splitting should therefore be understood within the resolved modal pair, with the launch-dependent power budget stated separately.

The complex modal overlap amplitudes $a_m(z)$ in figure 4 supply the complementary phase information. Their oscillation rate follows the 1S– $2P_y$ propagation-constant separation and is consequently faster for HC. Together, the predicted resonance, reciprocal power exchange, and phase evolution identify coherent modal coupling rather than an apparent redistribution caused by attenuation.

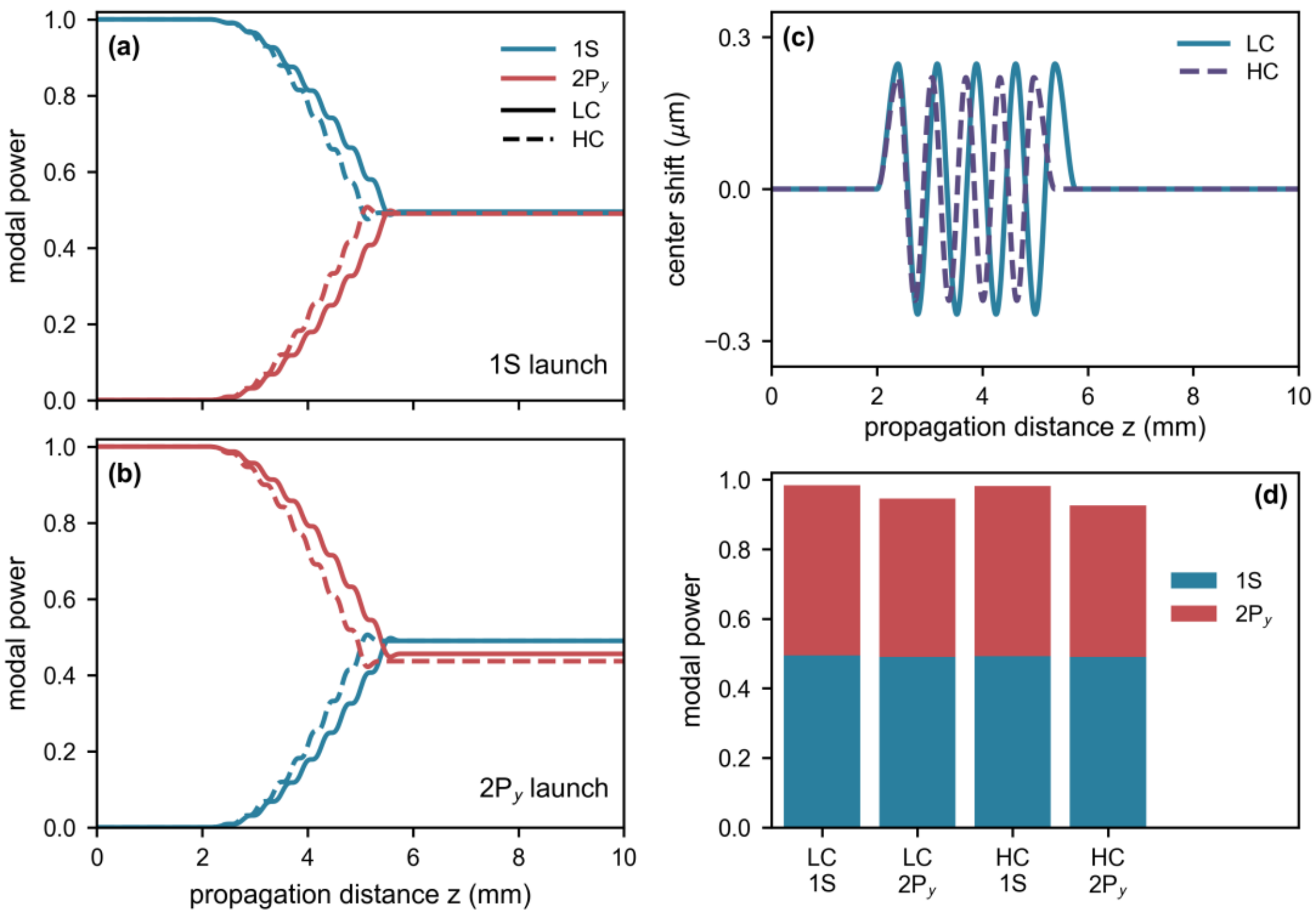

**Figure 3.** Reciprocal 1S↔2$P_y$ splitting in vertically modulated LC and HC waveguides. Panels (a) and (b) show absolute modal powers for 1S and 2$P_y$ launches; solid and dashed traces denote the LC and HC designs, respectively. Panel (c) shows the centerline trajectories: $A$ = 0.247 µm and $\Lambda$ = 0.744 mm for LC, and $A$ = 0.220 µm and Λ = 0.644 mm for HC. Both trajectories contain four full-amplitude periods and 0.40 mm cosine ramps. Panel (d) gives endpoint power in the resolved 1S/2$P_y$ pair; residual and absorber-removed power are not included in the bars.

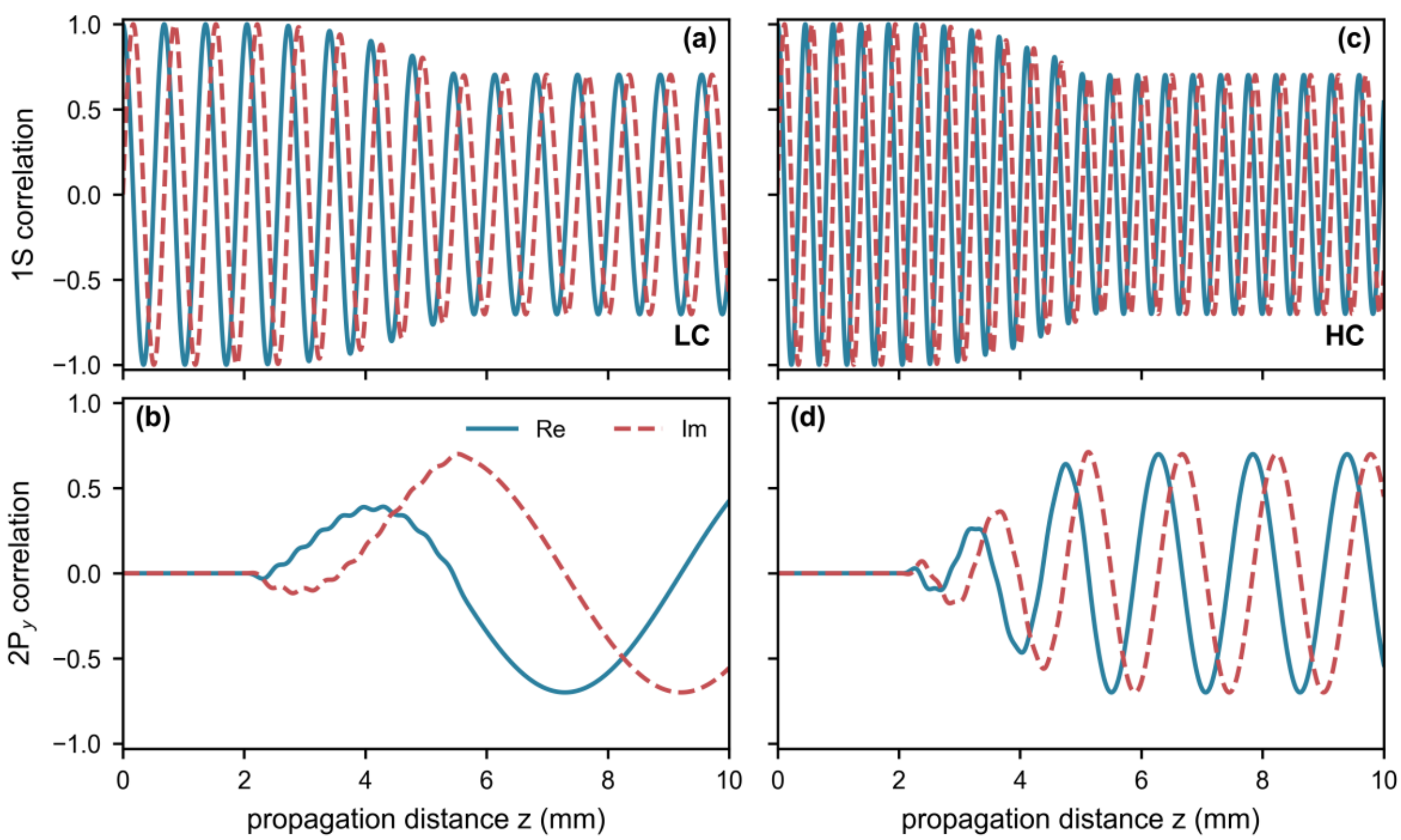

**Figure 4.** Complex modal overlap amplitudes for the same 1S-launched splitters. Panels (a,b) show the LC design and panels (c,d) the HC design; the upper and lower rows give overlaps with 1S and $2P_y$, respectively. Blue solid and red dashed curves are the real and imaginary parts. The faster HC phase evolution follows its larger 1S–$2P_y$ propagation-constant separation.

### 4.2 Parameter sensitivity

The usefulness of the modal splitter depends on two distinct error classes. Variations in ellipticity, core width, or index contrast alter the modal fields and their propagation-constant mismatch, whereas trajectory errors alter the phase-matching period or accumulated coupling strength. Figures 5 and 6 separate these mechanisms for the LC and HC designs on their physical parameter axes.

For computational efficiency, each perturbed-guide output is projected onto the validated basis of its nominal design rather than onto modes re-extracted at every sweep point. The plotted powers therefore measure fidelity to the intended device basis and include detuning, field mismatch, and propagation loss. They should not be interpreted as eigenvalue maps of the perturbed cross-sections.

The waveguide-parameter sweeps show distinct responses within the selected parameter intervals (figure 5). Core-width errors produce the flattest response over the calculated intervals, ellipticity produces a broader crossing around the nominal design, and index contrast most strongly detunes the equal-power condition. The same qualitative pattern appears in LC and HC, although the absolute curves reflect their different confinement. Because the axes span different fractional deviations from nominal values, these plots do not define a normalized sensitivity ranking.

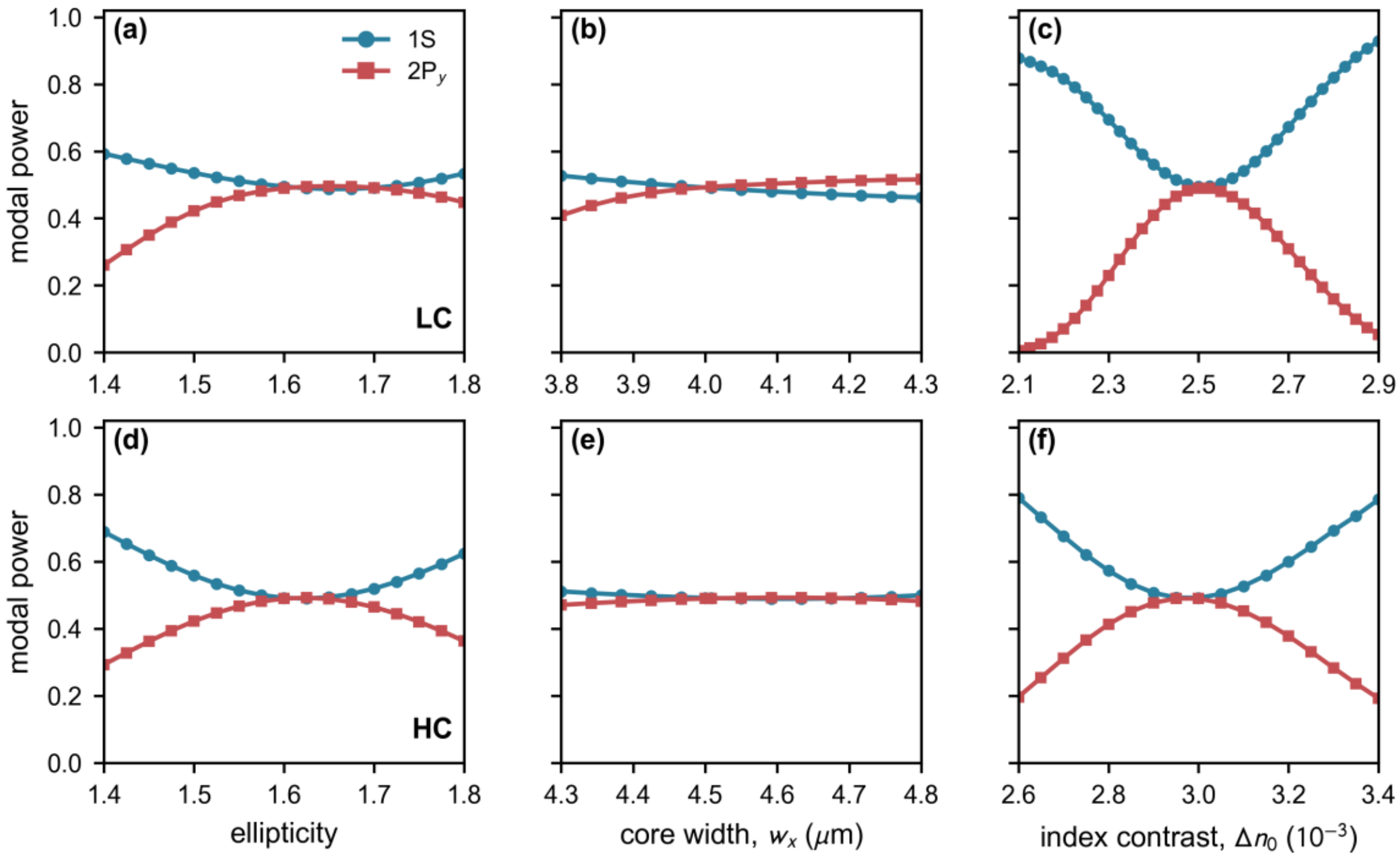

**Figure 5.** Sensitivity of the modal splitter to waveguide parameters for a validated 1S input. Columns show ellipticity, core width, and peak index contrast. Vertical axes scale is shared across the panels. The upper and lower rows show the LC and HC designs, respectively.

Trajectory period and amplitude affect the splitter in different ways (figure 6). Both LC and HC period scans show a finite phase-matching region around the selected value, whereas amplitude changes drive an approximately monotonic transfer through the equal-power point over the plotted interval. Within the parameter ranges examined here, amplitude variation produces a larger change in the output splitting ratio and therefore motivates exploration of post-fabrication trimming methods, such as thermo-optic perturbation. The latter is explored in section 6.

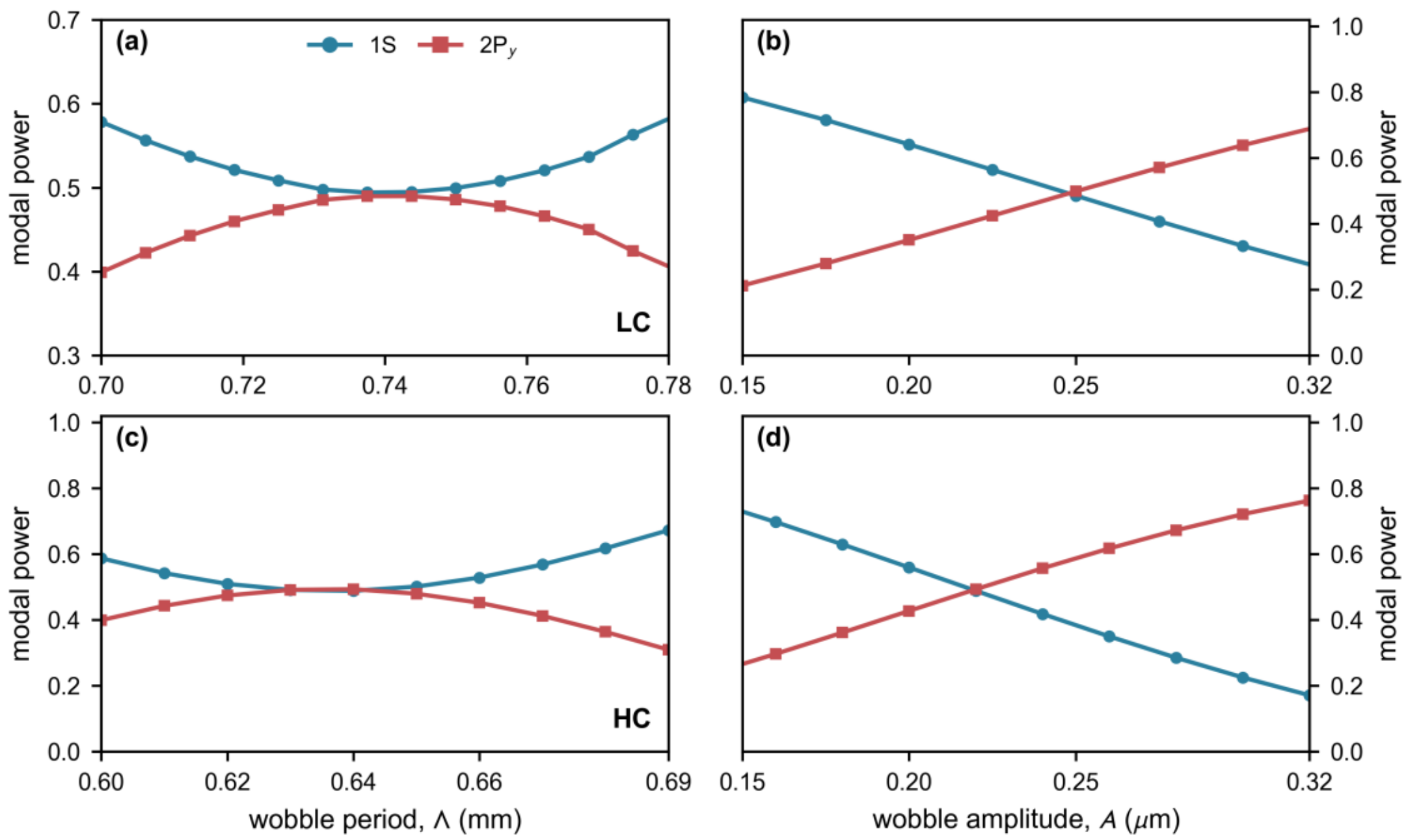

**Figure 6.** Sensitivity of the modal splitter to trajectory period and amplitude for a validated 1S input. The upper and lower rows show the LC and HC designs, respectively. The period scans use the selected amplitudes $A = 0.247$ μm (LC) and 0.220 μm (HC). The amplitude scans in panels (b) and (d) use $\Lambda = 0.744$ mm (LC) and 0.640 mm (HC). Markers denote calculated 1S and $2P_y$ powers.

## 5. Horizontal routing under the parity selection rule

Horizontal routing provides the complementary symmetry test. Unlike the vertical modulation used for conversion, the dominant perturbation of an S-bend in the $x - z$ plane is odd in $x$ and has the wrong $y$ parity for 1S↔$2P_y$ coupling in the ideal scalar guide. The reported displacement sweep uses a conformal mapping that keeps the guide on the numerical axis and represents curvature through an equivalent refractive-index profile,

$$n_{\mathrm{eq}}(x, y, z) = n(x, y)\exp[\kappa(z)x]. \quad (9)$$

The local curvature was calculated from the centerline $x_c(z)$ using:

$$\kappa(z) = \frac{x_c''(z)}{\left[1+(x_c'(z))^2\right]^{3/2}}. \quad (10)$$

The reported conformal calculations use a 4 mm cosine S-bend. The LC design is swept over 2–40 μm lateral displacement with a cleaned $2P_y$ input, and the HC design over 2–150 μm with separate cleaned $2P_y$ and 1S inputs. Each output is projected onto the modal basis. The HC calculations therefore track 1S, $2P_y$, and $2P_x$.

For a cleaned $2P_y$ input, retained modal power separates the effect of confinement from the parity rule (figure 7). LC falls to 0.814 at 40 μm displacement, whereas HC reaches 0.810 only at 150 μm. Across the HC sweep, transfer from $2P_y$ into 1S and $2P_x$ remains below order of $10^{-12}$ while the LC 1S projection is of order $10^{-13}$. These values coincide with the numerical modal-orthogonality floor. The unresolved HC output reaches 0.045 at 150 μm, showing that degradation of the $2P_y$ launch is radiation dominated within the model rather than conversion into the tracked modes.

The complementary HC 1S launch gives a different but equally selective response. At 150 μm displacement, 0.986 modal power remains in 1S, while $2P_x$ reaches 0.0060 and $2P_y$ remains below $3.5 \times 10^{-12}$. The total output power is 0.993 and the unresolved fraction is 0.0007 at this endpoint. The fundamental mode is therefore strongly retained, with the small resolved transfer confined to the symmetry-allowed $2P_x$ channel.

This pattern follows the first-order parity rule. The horizontal-bend perturbation is odd in $x$ and even in $y$, so it is mismatched to $1S\leftrightarrow 2P_y$ and $2P_y \leftrightarrow 2P_x$ but permits $1S\leftrightarrow 2P_x$. Despite being symmetry allowed, the latter transfer remains only 0.6% at the largest displacement because the modes are strongly detuned ($|\Delta\beta| \approx 1.31 \times 10^4$ rad m$^{-1}$) and the smooth trajectory is nearly adiabatic. This result does not imply immunity to arbitrary curvature or fabrication errors. Trajectory optimization can be applied in addition to the symmetry filter [6,13].

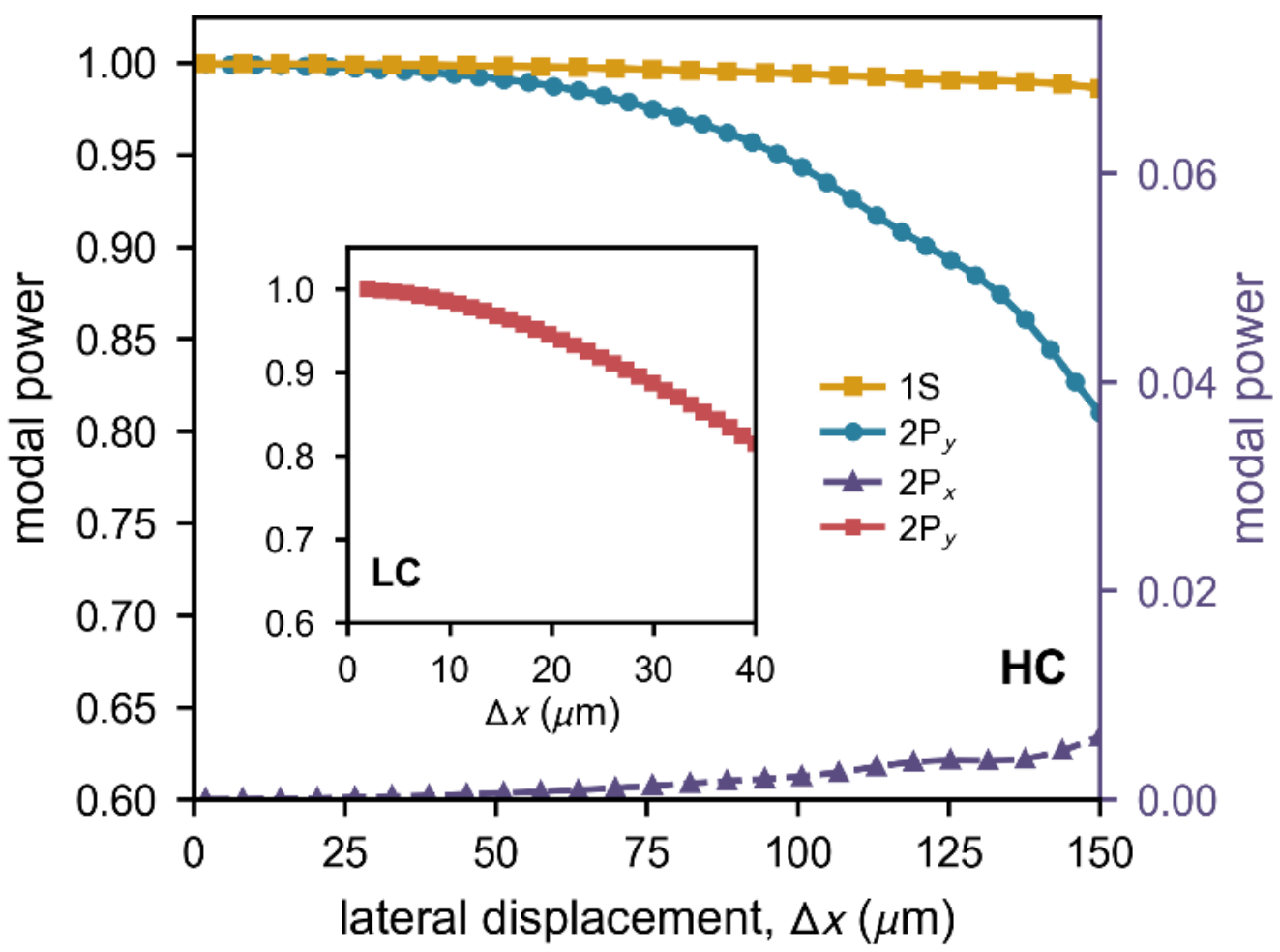

**Figure 7.** Modal powers after a 4 mm cosine S-bend. The main axes show HC retained 1S for a cleaned 1S input and retained $2P_y$ for a cleaned $2P_y$ input over 2–150 μm lateral displacement. Both refer to the left vertical axis. The $1S\rightarrow 2P_y$ and $2P_y\rightarrow 2P_x$ projections remain at the numerical floor and are omitted. The HC $1S\rightarrow 2P_x$ component refers to the right vertical axis. The inset shows LC retained $2P_y$ for a cleaned $2P_y$ input over 2–40 μm. Markers are calculated points, and connecting lines are guides to the eye.

## 6. Thermo-optic response and trimming

The first-order parity rule does not prevent a heater-induced index perturbation from detuning the modal splitter or increasing radiation. We therefore compare a surface-heater-like perturbation applied to the vertically modulated section with an equal-length straight-guide control [14]. In the straight guide the same perturbation tests unintended mode mixing, while over the splitter it tests post-fabrication adjustment of the $1S/2P_y$ balance.

To model this effect, the BPM simulations included an added thermo-optic index perturbation $\Delta n_{\text{th}}$,

$$n(x, y, z) = n_{\text{clad}} + \Delta n_{\text{wg}}(x, y, z) + \Delta n_{\text{th}}(x, y, z), \tag{11}$$

$$\Delta n_{\text{th}}(x, y, z; P) = \frac{P}{P_{2\pi}} \Delta n_{\text{th}}^{2\pi}(x, y) W(z). \tag{11a}$$

Equation (11a) defines the linear drive scaling used in the simulation. The transverse reference profile is the imposed surface-localized distribution shown in figure 8(d), and the longitudinal window is unity from 1.8 to 6.0 mm and zero elsewhere. Normalized heater drive, rather than absolute electrical power, is the simulation parameter. The conversion to milliwatts uses the illustrative 39 mW full-phase-shift value reported in reference 16. The imposed thermal index profile is surface localized rather than obtained from a full heat-equation solution. A heater with 10 µm transverse FWHM lies 30 µm above the guide and extends from $z$ = 1.8 to 6.0 mm. The drive is swept from 0 to 160 mW in 10 mW steps using $P_{2\pi}$ = 39 mW as a reference scale, while each LC and HC splitter retains its independently optimized trajectory. The $2P_y$ calculations use input and projection modes extracted independently on the 120 µm/768² grid.

The straight controls remain mode preserving, whereas a heater over the splitter changes the resolved output balance (figure 8). For a 1S launch, LC is more strongly detuned over the applied range than HC. A $2P_y$ launch reveals the accompanying confinement dependence: at zero drive the LC splitter gives 0.458 retained $2P_y$ and 0.493 converted 1S, while HC gives 0.435 and 0.490. These absolute powers sum to less than unity because unresolved and radiatively removed power is retained in the budget rather than hidden by pair renormalization. At 160 mW, HC remains close to an equal resolved split (0.468 and 0.466), whereas LC shifts to 0.368 and 0.420 and carries the larger unresolved-loss fraction. This indicates a confinement-dependent response, whereby HC is less sensitive to the imposed thermal profile, while LC provides a larger trimming range at the cost of greater unresolved and absorber-removed fractions.

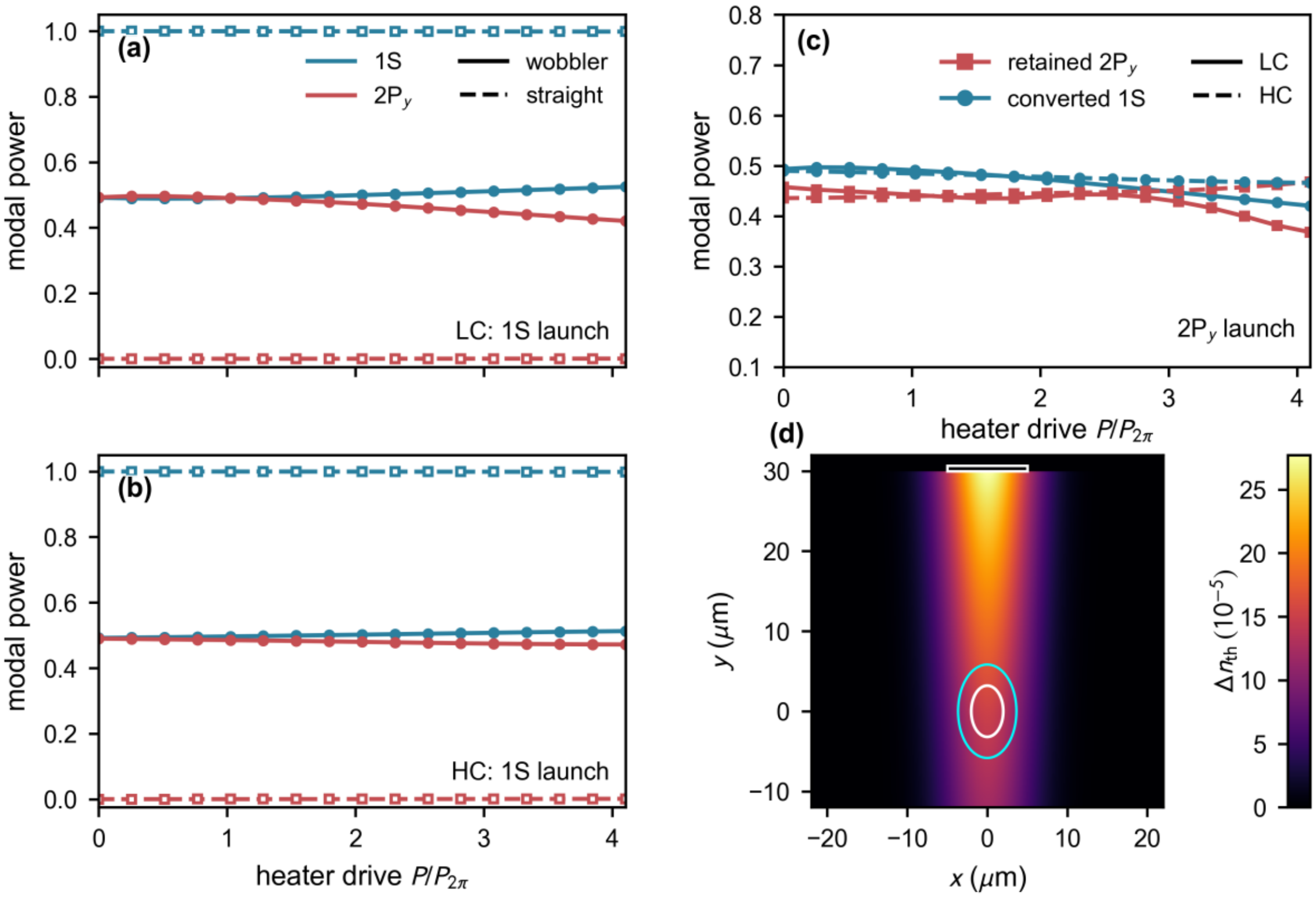

**Figure 8.** Thermo-optic response of the modal splitter. Panels (a) and (b) compare 1S-launched LC and HC splitters with straight-guide controls. Panel (c) shows absolute retained $2P_y$ and converted 1S powers for $2P_y$-launched splitters, using independently $\beta$-filtered modes. Zero drive denotes the unheated splitter, and the traces are not renormalized to the resolved modal pair. Panel (d) shows the imposed heater-induced index contrast, with the LC core outlined for scale.

## 7. Path–mode architecture enabled by the modal basis

The symmetry-selective waveguide can serve as a building block for integrated path–mode circuits [15–17]. Figure 9 combines an input horizontal directional coupler, independent vertically modulated sections, surface trim heaters, and an output coupler in a representative three-dimensional layout. This is a component-level architecture rather than an end-to-end simulation. The purpose is to show how routing and modal conversion can be assigned to orthogonal perturbation symmetries within one circuit.

For paths $|u\rangle$ and $|l\rangle$ and retained modes $|1S\rangle$ and $|2P_y\rangle$, a single-photon state in the resulting four-dimensional subspace can be written as

$$|\psi\rangle = c_{u,S}|u\rangle \otimes |1S\rangle + c_{u,P}|u\rangle \otimes |2P_y\rangle + c_{l,S}|l\rangle \otimes |1S\rangle + c_{l,P}|l\rangle \otimes |2P_y\rangle, \quad (12)$$

with normalization $\sum_{p,m}|c_{p,m}|^2 = 1$ over $p \in \{u, l\}$ and $m \in \{S, P_y\}$. In this basis the first horizontal coupler prepares a path superposition, while each vertically modulated section acts locally on the modal degree of freedom. An ideal 50:50 path coupler followed by independently tuned modal splitters and a path phase shift then gives states of the form

(13)

$$|\psi\rangle = \frac{1}{\sqrt{2}}\left[|u\rangle \otimes \left(\cos\theta_u|1S\rangle + e^{i\phi_u}\sin\theta_u|2P_y\rangle\right) + e^{i\delta}|l\rangle \otimes \left(\cos\theta_l|1S\rangle + e^{i\phi_l}\sin\theta_l|2P_y\rangle\right)\right].$$

Here $\theta_u$ and $\theta_l$ are set by the effective modal-coupling strengths, $\phi_u$ and $\phi_l$ are modal phases, and $\delta$ is the relative path phase. Equal splitting in both arms corresponds to $\theta_u = \theta_l = \pi/4$. In this ideal component picture, the output coupler recombines the path degree of freedom and the circuit acts as a path interferometer with embedded modal operations.

A heater overlapping a vertically modulated section can adjust its final 1S/$2P_y$ balance by changing the effective detuning, as in figure 8. A straight-section heater could instead trim relative modal phase only if its thermal field overlaps the two modes differently. This requires a transverse gradient or asymmetric placement. Any path–mode circuit inherits the symmetry sensitivity of these modal sections, so we next test whether longitudinal writing errors preserve or break the intended parity selection.

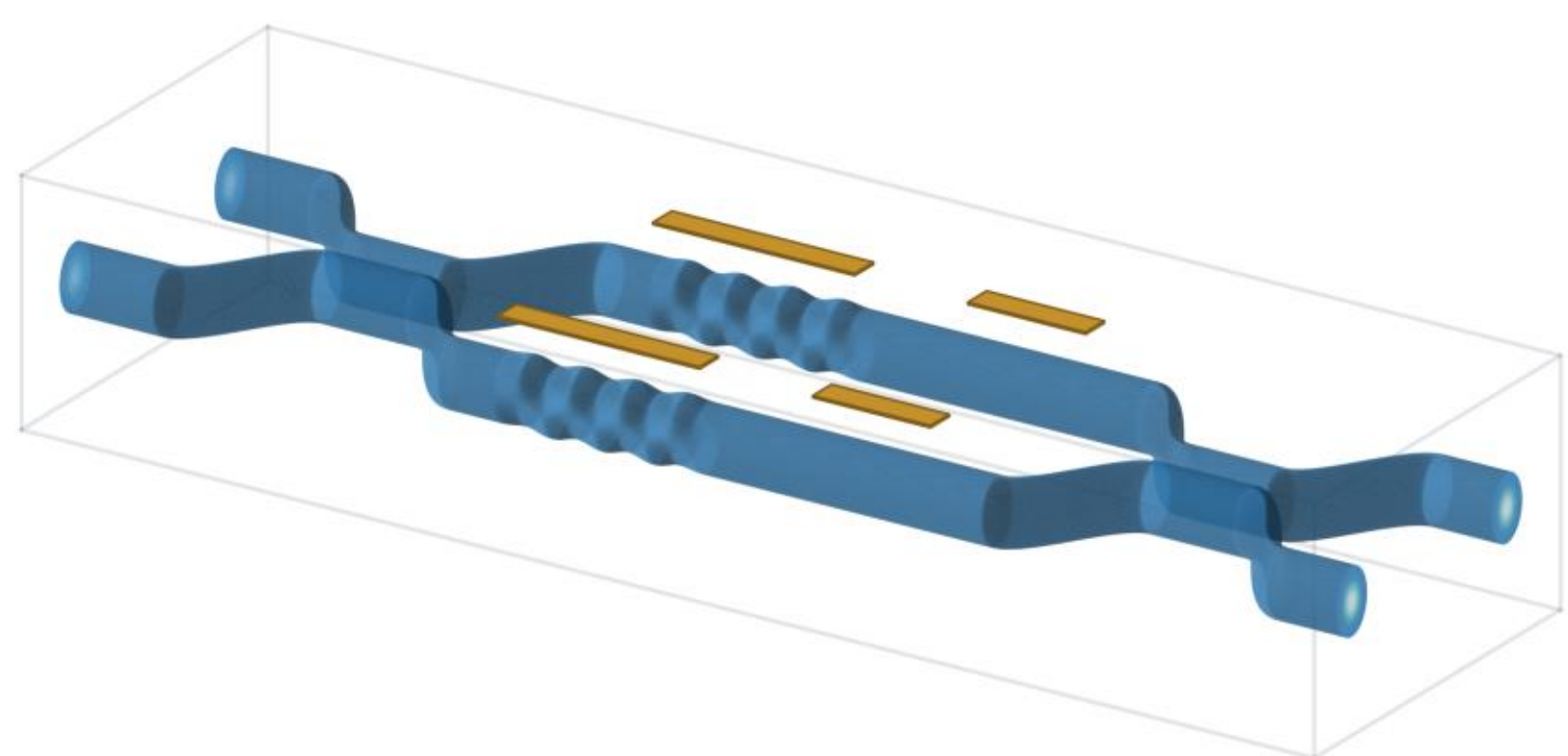

**Figure 9.** Conceptual three-dimensional path–mode architecture comprising input and output path couplers, vertically modulated modal-splitting sections, and surface thermal trim heaters.

## 8. Sensitivity to stochastic writing position errors

Writing-position noise can either preserve or break the parity underlying modal selectivity, depending on its direction. We therefore propagate the 1S mode through a straight 10 mm guide with stochastic longitudinal centerline displacement. Short-correlation vertical noise represents beam pointing, longer-correlation vertical noise represents servo-loop residuals, and vectorial $x - y$ noise tests explicit $x$-parity breaking. The LC output is projected onto its validated 1S/$2P_y$ basis and the HC output additionally onto guided $2P_x$.

The beam-pointing scan uses vertical root-mean-square (RMS) displacement $y$-RMS = 0–0.10 µm, correlation length 0.02 mm, and 25 realizations per point. The servo-loop scan uses $y$-RMS = 0–0.15 µm, correlation length 0.20 mm, and 25 realizations per point. The vectorial scan uses $y$-RMS = 0–0.15 µm, fixed $x$-RMS = 0.05 µm, correlation length 0.05 mm, and 15 realizations per point. Writing-position errors were represented by stationary Ornstein–Uhlenbeck centerline trajectories, updated at each propagation step and having exponential autocorrelation with the prescribed RMS displacement and the correlation length. Each ensemble member used a distinct deterministic random seed. Identical seed sets were used for the LC and HC designs to provide comparisons.

Vertical-only noise redistributes power within the intended 1S/$2P_y$ sector while preserving *x*-parity. At the largest beam-pointing error, mean 1S powers are 0.974 (LC) and 0.971 (HC), with $2P_y$ powers of 0.014 and 0.020. At the final servo-loop point *y*-RMS = 0.15 µm they become 0.920 (LC) and 0.895 (HC) in 1S, with corresponding 0.055 and 0.087 in $2P_y$. HC $2P_x$ remains at the $10^{-13}$ numerical floor. Small nonmonotonic variations lie within the finite-ensemble spread shown by the shaded bands. Adding an $x$ component activates the HC leakage channel (Fig. 10(f)). At *y*-RMS = 0.15 µm with *x*-RMS = 0.05 µm, LC gives mean powers $P_{1S}$ = 0.893 and $P_{2P_y}$ = 0.049. HC gives $P_{1S}$ = 0.852, $P_{2P_y}$ = 0.084, and $P_{2P_x}$ = 0.018. The vectorial noise, as expected, provides explicit $x$-parity breaking and leads to $1S - 2P_x$ coupling.

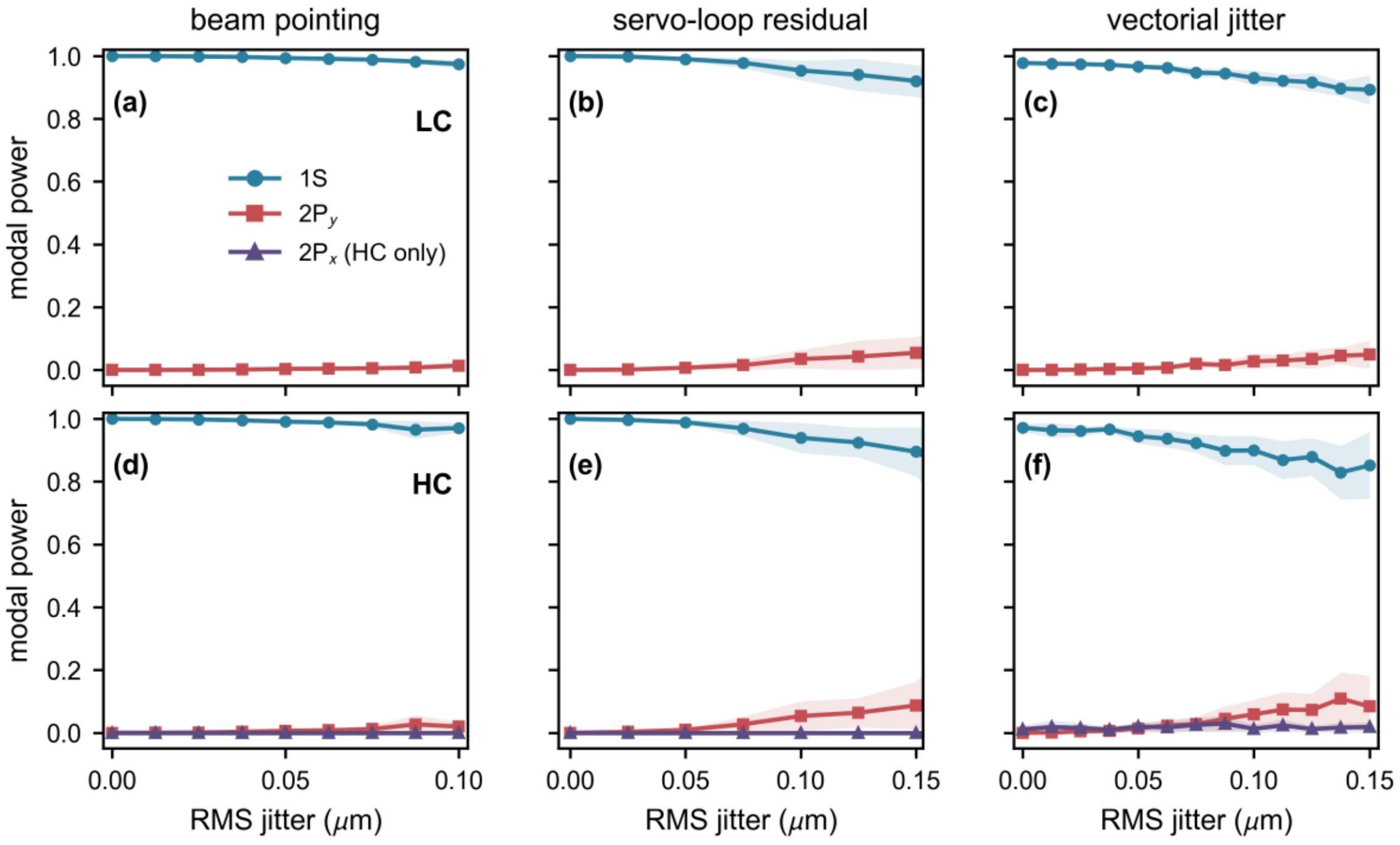

**Figure 10.** Sensitivity to stochastic writing-position errors after 10 mm of propagation in straight waveguide. Columns show beam-pointing, servo-loop, and vectorial position-noise models. The upper and lower rows show the LC and HC designs, respectively, on common vertical axes. Points and shaded bands give ensemble means and one standard deviation for a validated 1S input. The LC output is projected onto 1S and $2P_y$, whereas the HC output is also projected onto guided $2P_x$.

## 9. Discussion

The lower- and higher-confinement elliptical waveguide designs expose the central trade-off. LC design offers isolation of the $\{1S, 2P_y\}$ subspace, but its extended $2P_y$ mode is more susceptible to bend radiation. HC design improves retention yet supports a guided $2P_x$ mode. Vertical modulation and $y$-only jitter leave this mode at the numerical floor, while horizontal S-bending permits weak 1S→$2P_x$ transfer, and vectorial writing errors can also populate it. However, large 1S–$2P_x$ detuning keeps the bend-induced transfer small despite its allowed parity.

The calculations also distinguish modal crosstalk from radiation. The optimized vertical perturbation produces reciprocal, phase-matched 1S↔$2P_y$ exchange. For a $2P_y$ launch, horizontal S-bends produce no resolved transfer within numerical precision, although retained $2P_y$ decreases with curvature, particularly for LC. For the HC 1S launch, 0.986 remains in 1S at 150 μm and only 0.6% reaches $2P_x$. Symmetry therefore determines which coherent channels are allowed, while detuning and adiabaticity determine their strength. Neither mechanism prevents radiation from a strongly perturbed guide.

The path–mode architecture in figure 9 is an application outlook. Two paths carrying the $\{1S, 2P_y\}$ basis define a compact four-dimensional state space without a proportional increase in the number of written paths. Realizing that opportunity will require loss-balanced interfaces, phase-stable thermal control, and end-to-end modal readout. The present results support a numerical implementation of the modal selection rule and quantify its principal tolerance limits. An experimental implementation would additionally require path couplers whose splitting ratios are sufficiently mode independent, or independently calibrated for the 1S and $2P_y$ components.

## 10. Conclusion

Co-designing the supported modal spectrum and perturbation parity enables selective modal control in vertically elliptical femtosecond-laser-written waveguides. A periodic vertical displacement drives phase-matched 1S↔$2P_y$ exchange, whereas horizontal routing is parity mismatched to that transition at first order. For the HC design, the complementary 1S launch shows that the symmetry-allowed 1S→$2P_x$ channel reaches only 0.6% at the largest bend displacement, consistent with the large propagation-constant detuning and smooth routing. Agreement between modal-spectrum-predicted and optimized periods, reciprocal conversion, and coherent phase evolution supports this interpretation within the scalar BPM model. The LC/HC comparison reveals the practical design rule. No window-converged $2P_x$ leakage channel is resolved for LC, but its extended $2P_y$ mode is more vulnerable to radiation. Higher confinement improves retention under bending and the prescribed thermal perturbation, yet makes fidelity dependent on controlling $x$-parity-breaking errors that can populate guided $2P_x$. Confinement and symmetry must therefore be optimized together. Experimental implementation will require writing-position control to be characterized on the sub-100 nm scale. Experimental tests should first optimize the writing of core geometry and refractive-index contrast and validate the modal spectrum, polarization dependence, total-power budget, and thermal calibration of the proposed path–mode building block. This symmetry-filtered modal primitive could thus provide a compact route toward dense few-mode interferometers, modal beam splitters, spatial multiplexers, and three-dimensional quantum photonic circuits.

**Acknowledgments**

The authors acknowledge financial support from the Research Council of Lithuania, Project No. P-ITP-24-22.

**Data availability statement**

The data that support the findings of this study are available upon reasonable request from the authors.